# Properties of group-IV-based ferromagnetic semiconductor GeFe: Growth temperature dependence, lattice constant, location of Fe atoms, and their relevance to the magnetic properties

Yuki Wakabayashi, Shinobu Ohya, Yoshisuke Ban, and Masaaki Tanaka

*Department of Electrical Engineering and Information Systems, The University of Tokyo, 7-3-1 Hongo, Bunkyo-ku, Tokyo 113-8656, Japan*

We report the growth temperature dependence of the properties of the group-IV-based ferromagnetic semiconductor $Ge_{1-x}Fe_x$ films ($x$ = 6.5% and 10.5%), including the lattice constant, Curie temperature ($T_C$), and Fe-atom locations. While $T_C$ strongly depends on the growth temperature, we find a universal relationship between $T_C$ and the lattice constant, which does not depend on the Fe content $x$. By using the channeling Rutherford backscattering and particle-induced X-ray emission measurements, it is clarified that about 15% of the Fe atoms exist in the tetrahedral interstitial sites in the $Ge_{0.935}Fe_{0.065}$ lattice and that the substitutional Fe concentration is not correlated with $T_C$. Considering these results, we suggest that the non-uniformity of the Fe concentration plays an important role in determining the ferromagnetic properties of GeFe.

## I. INTRODUCTION

Group-IV-based ferromagnetic-semiconductor (FMS) GeFe is expected to be an efficient spin injector and detector which are compatible with Si- and Ge- based devices, because it can be epitaxially grown on Si and Ge substrates without a disordered interfacial layer or formation of ferromagnetic intermetallic Fe-Ge precipitates.[1]  One of the important characteristics of $Ge_{1-x}Fe_x$ is that the conductivity can be controlled by boron (B) doping independently of the Fe concentration $x$.[2]  Therefore, when a spin current is injected from GeFe to a nonmagnetic semiconductor, we can avoid the conductivity mismatch problem and suppress the spin flip scattering at the interfaces. However, the Curie temperature ($T_C$) of GeFe is currently at the highest 220 K and the detailed mechanism how it is determined has not been clarified yet.[3,4,5,6]  In this paper, we present the growth temperature ($T_S$) dependence of $T_C$ and the lattice constant of GeFe, that are investigated by magnetic circular dichroism (MCD) and the X-ray diffraction (XRD) measurements.  Also, we employ channeling Rutherford backscattering (c-RBS) and channeling particle-induced X-ray emission (c-PIXE) characterizations to investigate the location of the Fe atoms in the GeFe lattice.

## II. GROWTH

We have epitaxially grown $Ge_{1-x}Fe_x$ thin films with the Fe concentration $x$ of 0.065 and 0.105 on Ge(001) substrates by low-temperature molecular beam epitaxy (LT-MBE).  Figure 1(a) shows the schematic structure of the samples.  The growth process is described as follows.  After the Ge(001) substrate was chemically cleaned and its surface was hydrogen-terminated by buffered HF solution, it was introduced in the MBE growth chamber through an oil-free load-lock system.  After degassing the substrate at 400°C for 30 minutes and successive thermal cleaning at 900ºC for 15 min, we grew a 30-nm-thick Ge buffer layer at 200ºC, which was followed by the growth of a 120-nm-thick $Ge_{0.935}Fe_{0.065}$ layer at $T_S$ = 160 - 280ºC (8 samples) or a 120-nm-thick $Ge_{0.895}Fe_{0.105}$ layer at $T_S$ = 200 - 280ºC (6 samples).  After that, we grew a 2-nm-thick Ge capping layer at 200ºC to avoid the surface oxidation of the GeFe layer.  We used *in-situ* reflection high-energy electron diffraction (RHEED) to monitor the crystallinity and surface morphology of the Ge buffer layer, GeFe layer, and Ge capping layer during the growth.  Figure 1(b)-(d) shows the *in-situ* reflection high-energy electron diffraction (RHEED) patterns of (b) the Ge buffer layer surface, (c) 120-nm-thick

$Ge_{0.935}Fe_{0.065}$ layer surface grown at $T_S$ = 240°C, and (d) 2-nm-thick Ge capping layer surface with the electron-beam azimuth along the <110> direction of the Ge(001) substrate. The diffraction pattern of the Ge buffer layer surface showed intense and sharp 2 × 2 streaks, which indicate a 2-dimensional growth mode and exhibit a diamond-type single crystal structure, and also the weak Kikuchi lines indicating good crystallinity. The GeFe layer surface showed intense and sharp 2 × 2 streaks but with no clear Kikuchi lines. The Kikuchi lines appeared again after the growth of the Ge capping layer.

**III. MAGNETIC PROPERTIES**

We have carried out magneto-optical measurements on the GeFe films to investigate the $T_S$ dependence of $T_C$. MCD, which is defined as the difference between the optical reflectances of right- and left- circular polarized lights, is a powerful tool to investigate the magnetic properties of FMSs.[7] This is because the MCD intensity is proportional to the $s,p$-$d$ exchange interaction, which is considered to be the origin of the ferromagnetism in FMSs, and to the vertical component of the magnetization ($M$) in FMSs. Figure 2(a) shows the MCD spectra of the Ge substrate (blue curve), of the $Ge_{0.935}Fe_{0.065}$ films grown at $T_S$ = 160ºC (violet curve) and 240ºC (green curve), and of the $Ge_{0.895}Fe_{0.105}$ films grown at $T_S$ = 200ºC (yellow curve) and 240ºC (red curve), with a magnetic field $B$ of 1 T applied perpendicular to the film plane at 5 K. All the GeFe samples show the $E_1$ peak at around 2.3 eV corresponding to the $L$ point of the bulk Ge as we can see in the MCD spectrum of the Ge substrate. These $E_1$ peaks, that are enhanced by the $s,p$-$d$ exchange interaction in all the GeFe films, are the characteristic property of FMSs.[8] The broad peak ($E^*$) at around 1.4 eV observed in all the GeFe samples is thought to be related to the Fe-related impurity bands or d-d transitions.[6] Figures 2(b)–(d) show the $B$ dependence of the MCD intensities measured at 5 K with the incident photon energies of the $E^*$ (1.5 eV, black curve) and $E_1$ (2.3 eV, red dotted curve) for the $Ge_{0.935}Fe_{0.065}$ films grown at (b) 160ºC, (c) 240ºC, and (d) 280ºC. Figure 2(c) also shows the $B$ dependence of the normalized -$M$ (green dotted curve) measured by a superconducting quantum interference device (SQUID). Here, the diamagnetic signal of the Ge substrate was subtracted from the raw $M$ data. As shown in Fig. 2(b) and (c), in the $Ge_{0.935}Fe_{0.065}$ films grown at 160ºC and 240ºC, the shapes of the MCD-$B$ curves at 1.5 and 2.3 eV are identical with each other, which means that the MCD signals at 1.5 eV and 2.3 eV originate from the same ferromagnetic phase of GeFe.[9] Moreover, Fig. 2(c) shows that the shapes of the MCD-$B$ curves are the same as that of the -$M$ vs. $B$ curve measured by SQUID. This indicates that the $M$ data measured by SQUID has the same origin as that induces the spin splitting of the energy band of GeFe. These results indicate that the origin of the ferromagnetism is the single FMS phase of GeFe. In the $Ge_{0.935}Fe_{0.065}$ films grown at 280ºC, the shapes of the MCD-$B$ curves at 1.5 and 2.3 eV are not identical with each other (Fig. 2(d)), indicating that there are two or more magnetic phases in the film. From the same analyses on other samples, we have found that the $Ge_{1-x}Fe_x$ films grown with the range of $T_S$ from 160 to 260ºC have a single FMS phase for both of $x$ = 0.065 and 0.105.

Figure 3(a) shows the magnetic-field $B$ dependence of the MCD intensity of the $Ge_{0.935}Fe_{0.065}$ film grown at $T_S$ = 240°C measured with the incident photon energy of 2.3 eV at 5 K (blue curve), 80 K (violet curve), 100 K (yellow curve), and 110 K (red curve). The inset shows the close-up view near zero magnetic field. In Fig. 3(b), we have estimated $T_C$ by using the Arrott plots ($MCD^2$ - $B$/MCD) of the MCD-$B$ curves measured with the incident photon energy of 2.3 eV at various temperatures for the $Ge_{0.935}Fe_{0.065}$ film grown at $T_S$ = 240°C. In the plots, we can estimate the square of the spontaneous MCD ($\propto M$) by extrapolating the data in the high magnetic field. This method is well-established and convenient because it is free from the effect of the magnetic anisotropy which affects the low-magnetic-field properties and sometimes makes the accurate estimation of $T_C$ difficult. The estimated $T_C$ is 100 K in this film, and we can see that the same $T_C$ value is obtained both in the hysteresis loop analysis in Fig. 3(a) and the Arrott plots in Fig. 3(b).

Figure 4(a) shows the $T_S$ dependence of $T_C$ of the $Ge_{0.935}Fe_{0.065}$ ($x$=0.065) films grown at $T_S$ = 160 - 260ºC (blue points) and the $Ge_{0.895}Fe_{0.105}$ ($x$=0.105) films grown at $T_S$ = 200 - 260ºC (red squares). When $T_S \geq$ 280ºC (gray area), the GeFe films were phase-separated magnetically as mentioned above. Here, we estimated the $T_C$ values by using the Arrott plot of MCD at 2.3 eV. For both of $x$, the maximum $T_C$ is achieved when $T_S$ = 240ºC. These maximum $T_C$ values (100 K in $Ge_{0.935}Fe_{0.065}$ and 170 K in $Ge_{0.895}Fe_{0.105}$) are about 1.4 times higher than those in the previous study[5] (70 K in $Ge_{0.94}Fe_{0.06}$ and 120 K in $Ge_{0.905}Fe_{0.095}$). The saturation magnetization $M_S$, that was obtained by SQUID with a magnetic field of 1 T applied perpendicular to the film plane, in the $Ge_{0.935}Fe_{0.065}$ films grown at $T_S$ = 160ºC ($T_C$ = 20 K) and $T_S$ = 240ºC ($T_C$ = 100 K) is ~0.7$\mu_B$ per one Fe atom and ~1.3$\mu_B$ per one Fe atom, respectively, where $\mu_B$ is the Bohr magneton. These different $M_S$ values indicate that the density of the effective Fe atoms which contribute to the ferromagnetic ordering in the Ge$_{0.935}$Fe$_{0.065}$ film grown at $T_S$ = 240ºC is about twice higher than that in the Ge$_{0.935}$Fe$_{0.065}$ film grown at $T_S$ = 160ºC.

**IV. CRYSTALLOGRAPHIC ANALYSES**

Figure 5 shows (a) the XRD $\theta$-$2\theta$ spectrum and (b) the XRD rocking curve of the GeFe (004) reflection of the Ge$_{0.935}$Fe$_{0.065}$ film grown at $T_S$ = 240ºC. In Fig. 5(a), the (004) diffraction peak of the GeFe film is clearly seen on the higher-angle side of the Ge(004) peak with the clear fringes, which indicates that the film is a high-quality single crystal with an abrupt and smooth interface. The excellent crystallinity of the Ge$_{0.935}$Fe$_{0.065}$ film grown at $T_S$ = 240ºC is also confirmed by the XRD rocking curve of the GeFe(004) reflection shown in Fig. 5(b) exhibiting a narrow full width at half maximum of 0.03º, which is comparable to that of the Ge substrate, 0.01º.[10,11,12] Figure 4(b) shows the $T_S$ dependence of the lattice constant that is estimated from the XRD spectra of the Ge$_{0.935}$Fe$_{0.065}$ ($x$=0.065) films grown at $T_S$ = 160 - 260ºC (blue points) and the Ge$_{0.895}$Fe$_{0.105}$ ($x$=0.105) films grown at $T_S$ = 200 - 260ºC (red squares). For both of $x$, the lattice constant is minimum at $T_S$ = 240ºC. Figure 6 shows the lattice constant of the Ge$_{0.935}$Fe$_{0.065}$ films grown at $T_S$ = 160 - 260ºC (blue points) and the Ge$_{0.895}$Fe$_{0.105}$ films grown at $T_S$ = 200 - 260ºC (red squares), plotted as a function of $T_C$. We see a universal relationship between the $T_C$ and the lattice constant, which does not depend on $x$. The $T_C$ increases as the lattice constant of the GeFe films decreases.

We employed c-RBS and c-PIXE to determine the position of the Fe atoms in the host Ge lattice. Figure 7 shows our experimental configurations of the PIXE Fe K$\alpha$ and the RBS angular scans in the {100} plane (a) around the <100> axis and (b) around the <110> axis. The red dotted arrows represent the direction of the incident $^4$He$^{++}$ beam. We measured the scattering yields as a function of the incident angle $\varphi$, that is defined as the angle between the crystal-axial (<100> or <110>) direction and the incident direction of an energetic ion beam of $^4$He$^{++}$, by varying $\varphi$ in a few degrees from 0º. Ideally, the scattering yields have a minimum value at $\varphi$=0º, where the incident beam is perfectly aligned in the crystal axis and the scattering is most suppressed, while they tend to increase and finally saturate when $|\varphi|$ becomes larger than several degrees.[13-16] We define $\chi^d(A)$ as the scattering yield originating from the element $A$ normalized by the averaged value of the scattering yields at $|\varphi|$ ranging from 2.5 to 3º when the measurement is carried out around the $d$ crystal-axial direction. Also, $\chi^d_{min}(A)$ is defined as the minimum value of $\chi^d(A)$, that is obtained at $\varphi=0°$. Figure 8(a) shows the schematic illustration of the possible sites that can be occupied in the diamond-type lattice structure, including the substitutional (or host) sites S (yellow spheres), the bond-center site BC (black sphere), the antibonding sites Q (violet spheres), the hexagonal site H (blue sphere), and the tetrahedral sites T (red spheres).[17,18] Here, we refer to these sites as specific sites. Figures 8(b) and 8(c) illustrate the locations of these specific sites when seen from the <100> and <110> directions, respectively. We note that other sites that are equivalent to these specific sites are neglected in Fig. 8 for simplicity.

Figure 9 shows the PIXE-Fe-Kα and RBS angular scans in the {100} plane around the <100> axis [(a) and (c)] and around the <110> axis [(b) and (d)] for the Ge$_{0.935}$Fe$_{0.065}$ films grown at (a) (b) $T_S$ = 160ºC and (c) (d) $T_S$ = 240ºC measured with a 2.275-MeV-$^4$He$^{++}$ beam. In all the graphs, the normalized yield has a minimum value at $\varphi \approx 0°$, which indicates that a lot of Fe atoms are shadowed by the host atoms and that they are not visible to the beam along the <100> or <110> axial directions. The important point is that $\chi^{<100>}_{min}$(Fe) is almost the same as $\chi^{<100>}_{min}$(Ge) in both samples [Fig. 9(a) and 9(c)]. This result indicates that most Fe atoms are located on the S or T sites because the specific sites which are shadowed by the host atoms in the <100> axial direction are only the S and T sites as we can see in Fig. 8(b). We define $f_{S+T}$ as the sum of the densities of the Fe atoms located on the S and T sites divided by the total density of the Fe atoms. We can roughly estimate $f_{S+T}$ by using the following equation with $\chi^{<100>}_{min}$(Fe) and $\chi^{<100>}_{min}$(Ge) obtained in the <100> axial direction:[11]

$$f_{S+T} = \left[1 - \chi^{<100>}_{min}(\text{Fe})\right] / \left[1 - \chi^{<100>}_{min}(\text{Ge})\right]. \tag{1}$$

By using eq. (1), $f_{S+T}$ in the Ge$_{0.935}$Fe$_{0.065}$ films grown at $T_S$ = 160ºC and $T_S$ = 240ºC is estimated to be 93% and 96%, respectively. These results mean that, in the Ge$_{0.935}$Fe$_{0.065}$ films grown at $T_S$ = 160ºC and $T_S$ = 240ºC, the small fraction of 7% and 4% of the doped Fe atoms is located on the sites other than the S or T sites. The second important point is that $\chi^{<110>}_{min}$(Fe) is higher than $\chi^{<110>}_{min}$(Ge) in the both Ge$_{0.935}$Fe$_{0.065}$ films grown at $T_S$ = 160ºC [Fig. 9(b)] and $T_S$ = 240ºC [Fig. 9(d)]. As can be seen in Figs. 8(b) and 8(c), the Fe atoms on the T sites in the diamond lattice are shadowed by the host atoms in the <100> direction but are exposed in the <110> direction. The atoms on the T sites are irradiated by the beam in all the $\varphi$ range shown in Fig. 9(b) and 9(d). It is known that the yields coming from the exposed atoms on the T sites are slightly enhanced in the vicinity of the <110> direction ($\varphi \approx 0°$) due to the flux peaking effect,[19] which has been well understood both theoretically[20] and experimentally.[21] The flux peaking effect is induced by the increase in the flux of the incident energetic ion beam near the T sites in the <110> direction.[17,18,22] From the numerical calculation, the enhancement factor $F$ for a surface layer, which is defined as the ratio of the intensity of the Fe-K$\alpha$-X-rays coming from the atoms on the T sites at $\varphi=0°$ to that at $|\varphi|=2°$, has been estimated to be in the range of 1.8 - 2.2 in the <110> axes for a diamond cubic lattice.[13,17] The ratio $f_T$ of the density of the Fe atoms on the T sites to the sum of those on the S and T sites can be roughly obtained from the <110> channeling results by the following equation:[23]

$$\chi_{\min}^{<110>}(\text{Fe}) = (1 - f_{S+T}) + f_{S+T}[F f_T + \chi_{\min}^{<110>}(\text{Ge})(1 - f_T)]. \qquad (2)$$

By eq. (2), $f_T$ in the Ge$_{0.935}$Fe$_{0.065}$ films grown at $T_S$ = 160ºC and $T_S$ = 240ºC is estimated to be 13-16% and 15-18%, respectively. In the case of typical III-V-based FMS GaMnAs, the similar fractions of the Mn atoms on the T sites have been found,[24,25] and it has been known that the density of the substitutional Mn atoms is correlated with $T_C$, which means that they determine the ferromagnetic properties of GaMnAs films.[26,27] However, in the case of GeFe, we do not see a clear difference in the Fe density on the T sites between the two samples, which have the large difference in the $T_C$ values (20 K and 100 K for the films grown at $T_S$ = 160ºC and $T_S$ = 240ºC, respectively). This indicates that the ferromagnetism in GeFe is not directly related to the density of the substitutional Fe atoms.

**V. DISCUSSION**

As described above, the $f_T$ values obtained in the Ge$_{0.935}$Fe$_{0.065}$ films grown at $T_S$ = 160ºC and $T_S$ = 240ºC are almost the same, but the lattice constants (5.654 and 5.647 Å, respectively) and $T_C$ values (20 and 100 K, respectively) are different. This difference in the lattice constant may be attributed to the difference in the density of the stacking-fault defects along the (111) plane observed in the GeFe films.[5,6] In face-centered cubic crystals, the (004) diffraction peak shifts to the higher angle side with an increase in the density of the stacking-fault defects along the (111) plane.[28,29,30]

We observe many stacking-fault defects especially in the locally high-Fe-concentration regions.[5] We think that the higher the $T_S$ is, the larger the non-uniformity of the Fe concentration becomes, which results in the increase in the density of the stacking-fault defects. Thus, the difference in the lattice constant may reflect the difference in the non-uniformity of the Fe concentration, which is thought to be the origin of the difference in the $M_S$ and $T_C$ values in the GeFe films.[31,32] When the non-uniformity of the Fe concentration is enhanced, the locally high-Fe-concentration regions may easily connect each other magnetically, resulting in higher $T_C$. In III-V and II-VI-based FMSs, such an increase in $T_C$ due to the enhancement of the non-uniformity of the magnetic impurities is predicted theoretically[32] by the random phase approximation in the Heisenberg model.[33,34] In fact, the $T_C$ value of the $Ge_{0.895}Fe_{0.105}$ film grown at $T_S$ = 240°C is increased from 170 K to 220 K by the post-growth annealing at 500°C due to the enhancement of the non-uniformity of the Fe concentration,[6] while a single ferromagnetic phase with the diamond-type crystal structure is kept in the film. We note that the GeFe films have a single FMS phase which is confirmed by the MCD and SQUID measurements as mentioned in Sec. III, even though they have the non-uniformity of the Fe concentration. This means that, once ferromagnetism appears, the locally high-Fe-concentration regions and low-Fe-concentration regions are magnetically coupled by the $s,p$-$d$ exchange interaction, which is confirmed by the enhancement of the $E_1$ peak in the MCD spectra (Fig. 2(a)).

## VI. SUMMARY

We have studied the $T_S$ dependence of the $Ge_{1-x}Fe_x$ films with $x$ = 0.065 and 0.105, and observed continuous changes in $T_C$ and in the lattice constant as a function of $T_S$. The $T_C$ value increases with the decrease in the lattice constant, and the relationship between $T_C$ and the lattice constant does not depend on $x$. The $T_C$ reaches maximum values of 100 and 170 K when $T_S$ = 240ºC for $x$ = 0.065 and 0.105, respectively. These maximum $T_C$ values are about 1.4 times higher than those grown at 200ºC in the previous study.[5] By using c-RBS and c-PIXE characterizations, we have found that the small fraction of 7% and 4% of the doped Fe atoms are located on the sites other than the S or T sites in the $Ge_{0.935}Fe_{0.065}$ samples with $T_C$ of 20 and 100 K, respectively. Among the Fe atoms located on the S and T sites, 13-16% and 15-18% of them exist on the T sites in the $Ge_{0.935}Fe_{0.065}$ films with the different $T_C$ values of 20 and 100 K, respectively. By considering these results, we suggest that the non-uniformity of the Fe concentration plays an important role in determining the ferromagnetic properties of the GeFe films. The influences of the tetrahedral interstitial magnetic impurities on the ferromagnetic properties are completely different between GeFe and GaMnAs.

**ACKNOWLEDGEMENTS**

This work was partly supported by Giant-in-Aids for Scientific Research including Specially Promoted Research, Project for Developing Innovation Systems of MEXT, and FIRST program of JSPS.

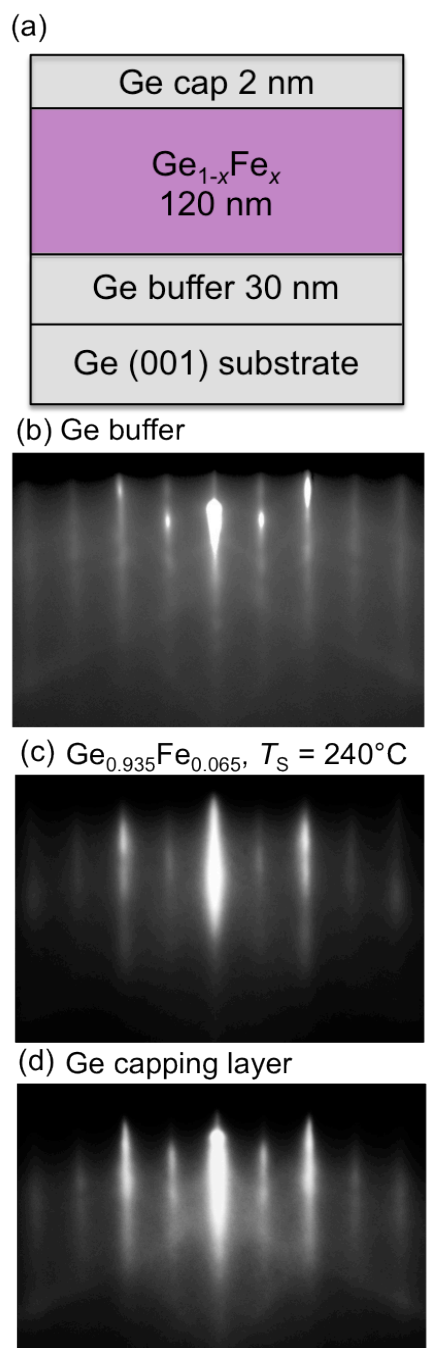

Fig. 1. (a) Schematic structure of the samples consisting of Ge cap (2 nm) / $Ge_{1-x}Fe_x$ (120 nm) / Ge buffer (30 nm) / Ge(001) substrate. (b)-(d) *In-situ* reflection high-energy electron diffraction (RHEED) patterns of (b) the Ge buffer layer surface, (c) 120-nm-thick $Ge_{0.935}Fe_{0.065}$ layer surface grown at $T_S$ = 240°C, and (d) 2-nm-thick Ge capping layer surface with the electron-beam azimuth along the <110> direction of the Ge (001) substrate.

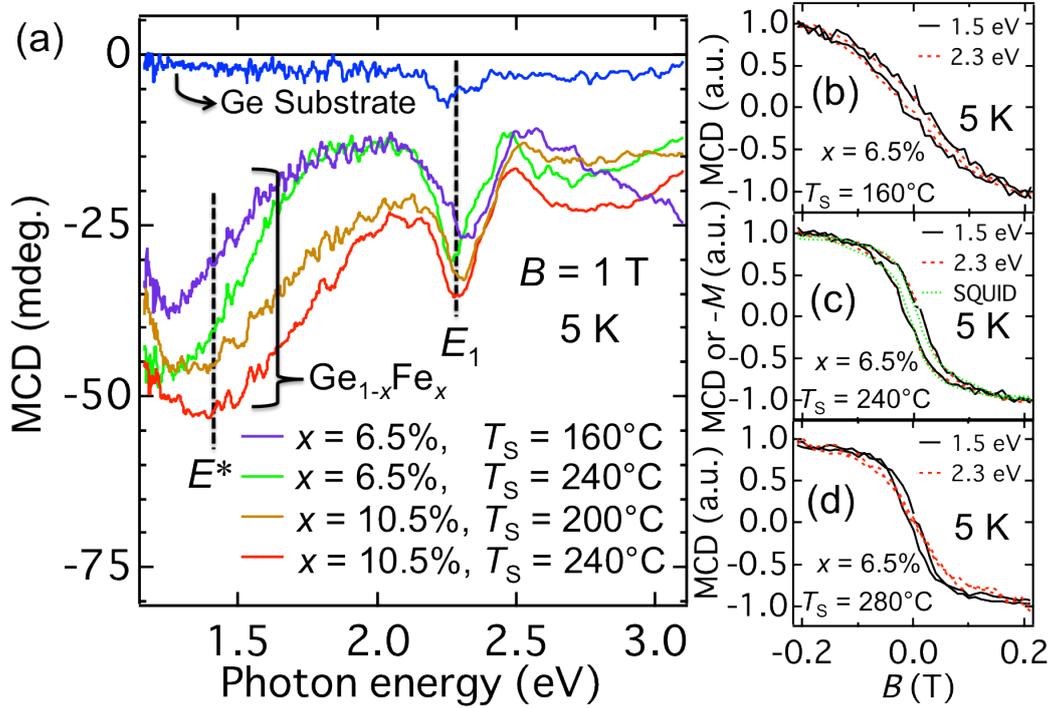

Fig. 2. (a) MCD spectra of the Ge substrate (blue curve), of the $Ge_{0.935}Fe_{0.065}$ ($x$ = 0.065) films grown at $T_S$ = 160ºC (violet curve) and $T_S$ = 240ºC (green curve), and of the $Ge_{0.895}Fe_{0.105}$ ($x$ = 0.105) films grown at $T_S$ = 200ºC (brown curve) and $T_S$ = 240ºC (red curve), with a magnetic field $B$ of 1 T applied perpendicular to the film plane at 5 K. (b)-(d) MCD intensity as a function of $B$ measured at 5 K with the photon energies of 1.5 eV (black curve) and 2.3 eV (red dotted curve) for the $Ge_{0.935}Fe_{0.065}$ films grown at (b) $T_S$ = 160ºC, (c) $T_S$ = 240ºC, and (d) $T_S$ = 280ºC. In (c), the green dotted curve expresses the $B$ dependence of -$M$ measured by SQUID.

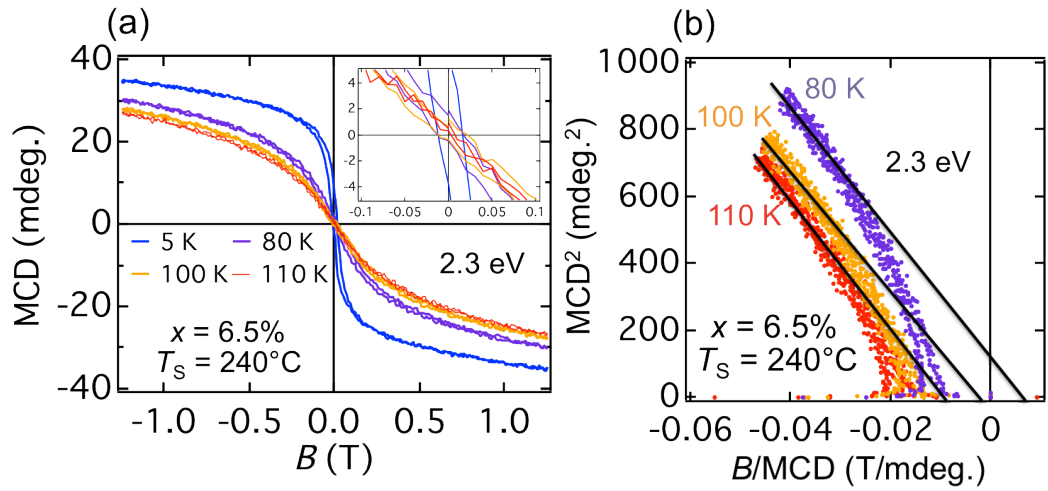

Fig. 3. (a) MCD intensity as a function of $B$ of the Ge$_{0.935}$Fe$_{0.065}$ film grown at $T_S$ = 240°C measured with the incident photon energy of 2.3 eV at 5 K (blue curve), 80 K (violet curve), 100 K (yellow curve), and 110 K (red curve). The inset shows the close-up view near zero magnetic field. (b) Arrott plots of the MCD-$B$ data measured with the incident photon energy of 2.3 eV at various temperatures for the Ge$_{0.935}$Fe$_{0.065}$ film grown at $T_S$ = 240°C.

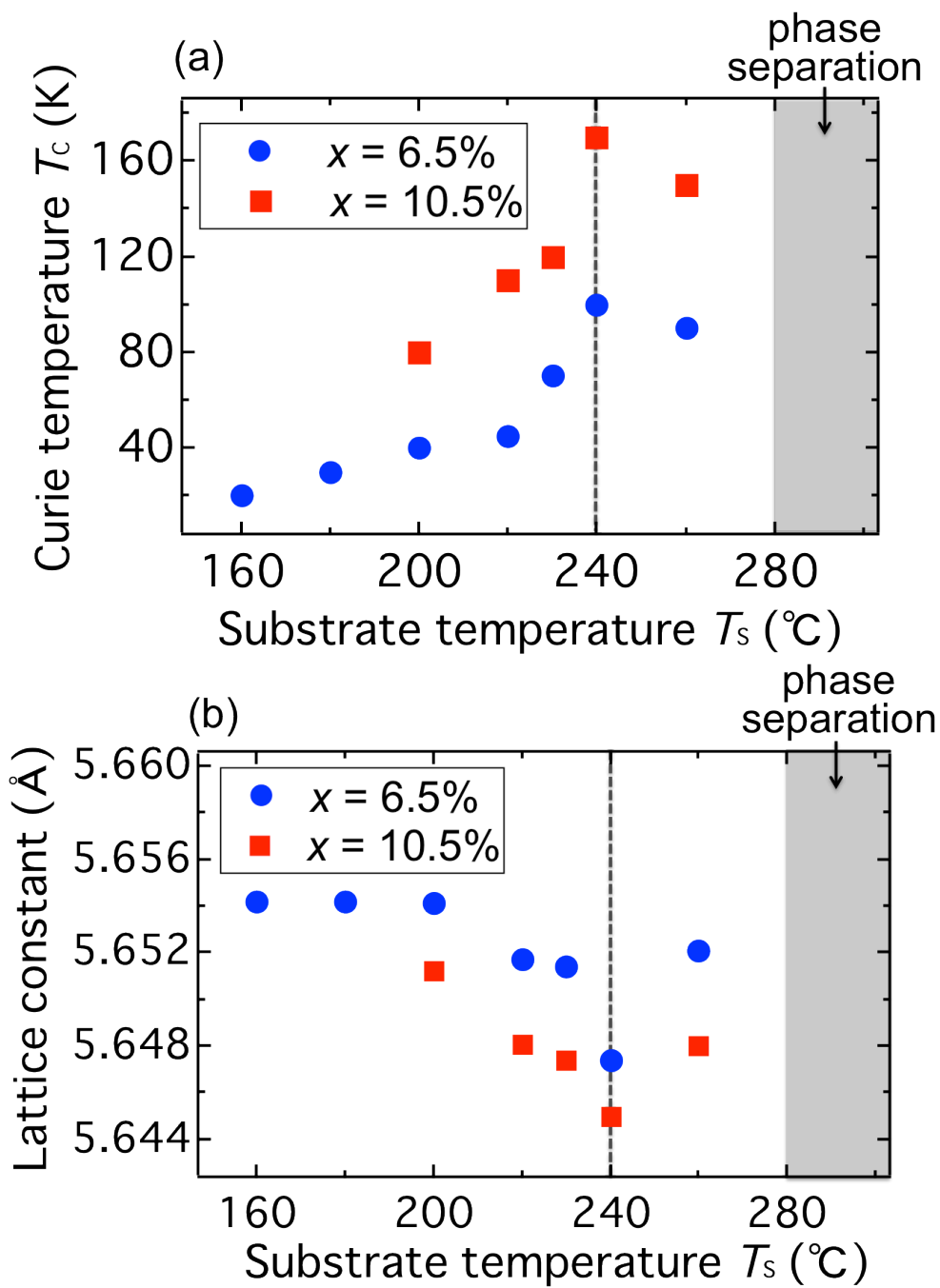

Fig. 4. (a) Curie temperature $T_C$ as a function of $T_S$ of the $Ge_{0.895}Fe_{0.105}$ films (blue points) and $Ge_{0.895}Fe_{0.105}$ films (red squares), that were estimated by the Arrott plots ($MCD^2$ - $B$/MCD) with the photon energy of 2.3 eV, where $B$ is the magnetic field. (b) Lattice constant as a function of $T_S$ of the $Ge_{0.935}Fe_{0.065}$ ($x$=0.065) films grown at $T_S$ = 160-260ºC (blue points) and the $Ge_{0.895}Fe_{0.105}$ ($x$=0.105) films grown at $T_S$ = 200-260ºC (red squares).

-

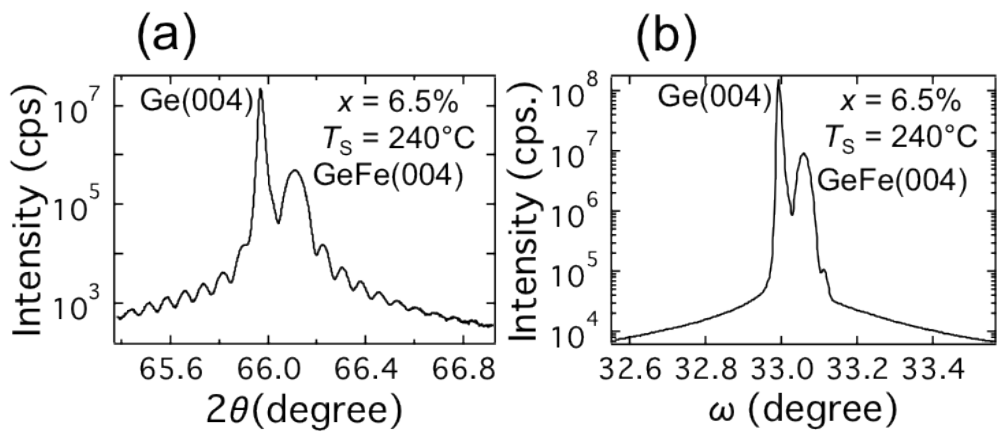

Fig. 5. XRD (a) $\theta$-$2\theta$ spectrum and (b) rocking curve of the GeFe (004) reflection of the Ge$_{0.935}$Fe$_{0.065}$ film grown at $T_S$ = 240°C.

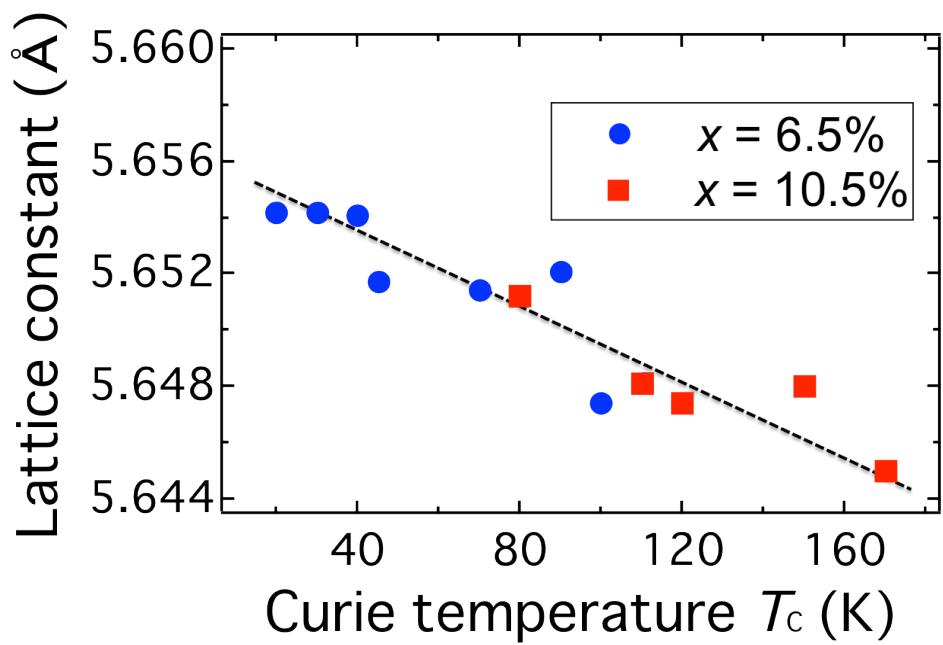

Fig. 6. Lattice constant estimated from the XRD spectra plotted as a function of $T_C$ in the Ge$_{0.895}$Fe$_{0.105}$ films (blue points) and the Ge$_{0.895}$Fe$_{0.105}$ films (red squares).

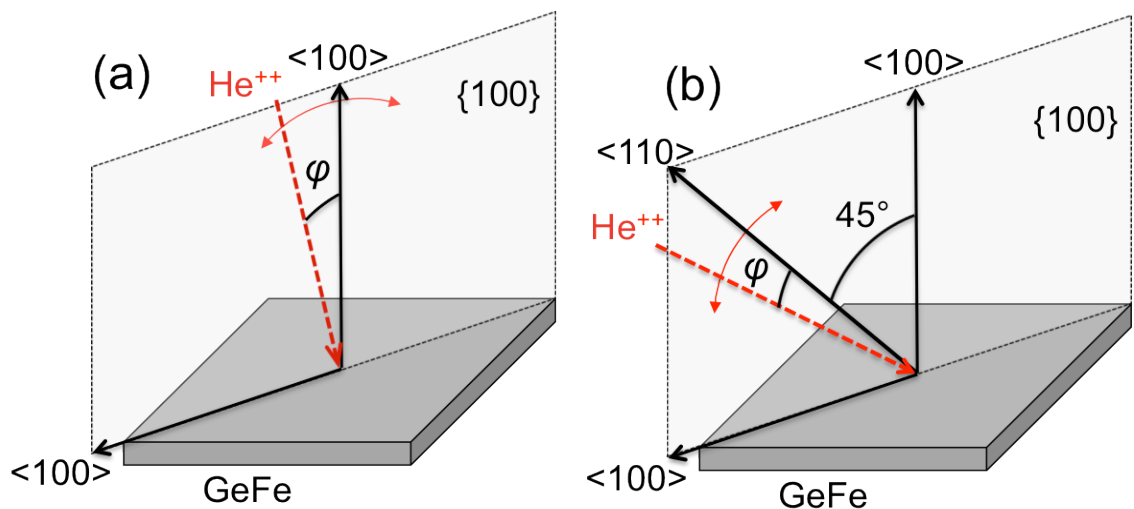

Fig. 7. Experimental configurations of the PIXE Fe Kα and RBS angular scans in the {100} plane (a) around the <100> axis and (b) around the <110> axis. The (red) dotted arrows represent the direction of the incident $^4$He$^{++}$ beam, and $\varphi$ represents the angle between the incident beam direction and the axial direction.

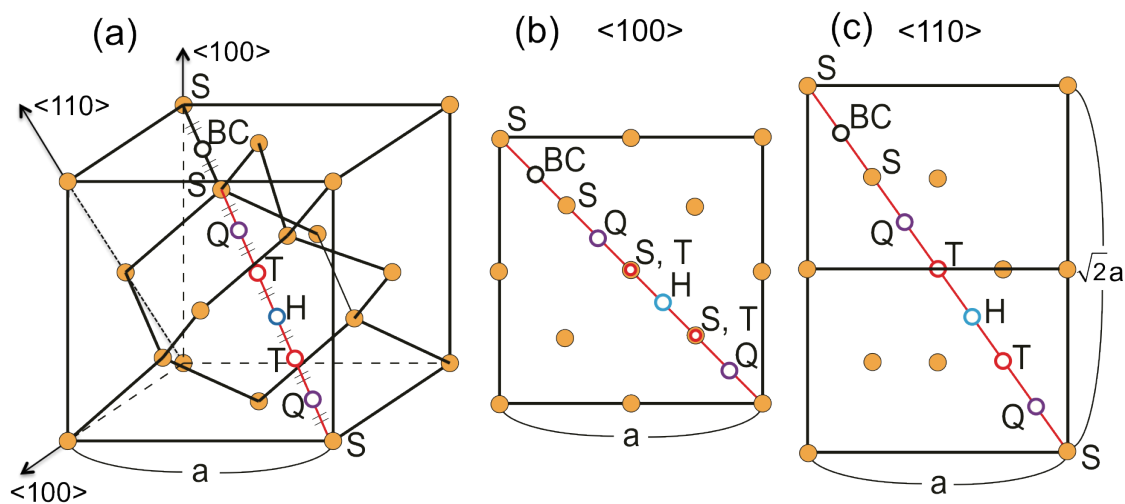

Fig. 8. (a) Schematic illustration of the possible specific sites in the diamond-type lattice structure, including the substitutional sites S (yellow spheres), the bond-center site BC (black sphere), the antibonding sites Q (violet spheres), the hexagonal site H (blue sphere), and the tetrahedral sites T (red spheres). (b),(c) The location of the specific sites when seen from the (b) <100> and (c) <110> directions.

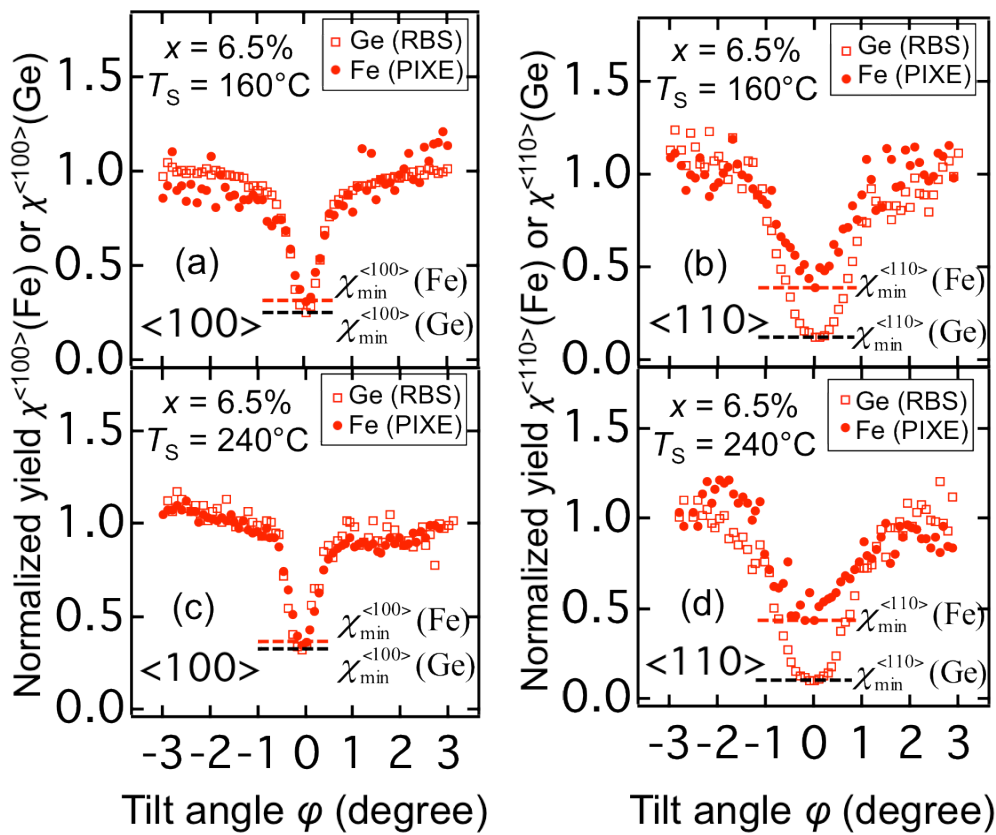

Fig. 9. PIXE-Fe-Kα (red points) and RBS (red squares) angular scans around (a) (c) the <100> axis and (b) (d) the <110> axis for the $Ge_{0.935}Fe_{0.065}$ films grown at (a) (b) $T_S$ = 160ºC and (c) (d) $T_S$ = 240ºC.